\documentclass[12pt,aps,floats]{revtex4}

\usepackage{graphicx}
\usepackage{epsfig}

\newcommand{\be}{\begin{equation}}
\newcommand{\ee}{\end{equation}}
\newcommand{\bea}{\begin{eqnarray}}
\newcommand{\eea}{\end{eqnarray}}

\def\simlt{\stackrel{<}{{}_\sim}}
\def\simgt{\stackrel{>}{{}_\sim}}
\begin{document}
\title{Spatial correlations of primordial density fluctuations in the
standard cosmological model}

\author{David H. Oaknin}

\affiliation{
Department of Physics, Technion, Haifa, 32000, Israel. \\
e-mail: doaknin@physics.technion.ac.il}

\begin{abstract}
We revisit the {\it origin of structures problem} of standard 
Friedmann-Robertson-Walker cosmology to point out an unjustified 
approximation in the prevalent analysis. We follow common procedures 
in statistical mechanics to revise the issue without the disputed 
approximation. Our conclusions contradict the current wisdom and reveal
and unexpected scenario for the origin of primordial cosmological structures. 
We show that standard physics operating in the cosmic plasma during 
the radiation dominated expansion of the universe produce at the time of
decoupling scale invariant density anisotropies over cosmologically large 
comoving volumes. Scale invariance is shown to be a direct consequence of the 
causality constrains imposed by the short FRW comoving horizon at decoupling, 
which strongly suppress the power spectrum of density fluctuations with 
cosmologically large comoving wavelength. The global amplitude of these 
cosmological density anisotropies is fixed by the power spectrum in comoving 
modes whose wavelength is shorter than the causal horizon at the time and can 
be comparable to the amplitude of the primordial cosmological inhomogeneities 
imprinted in the cosmic microwave background radiation.
\end{abstract}

\maketitle

\section{Introduction.}
The cosmic microwave background radiation~(CMBR) is a fossil evidence of the 
early history of the expanding and cooling Friedmann-Robertson-Walker~(FRW) 
universe. Indeed, this radiation is a high-resolution picture of 
the cosmological density field at the instant of recombination, when the 
temperature of the universe was $T_{eq} \sim 1$~eV and this radiation last 
scattered and decoupled from the cosmic plasma. At that instant the size of 
the universe was $10^{-3}$ of its current size. The red shifted radiation 
observed at present is extremely homogeneous and isotropic over the sky, 
with a perfect blackbody spectrum of temperature $T = 2.725$~K~$\sim 10^{-3}$~eV 
\cite{Smoot:1992td} up to tiny fluctuations $\Delta T/T \simlt 10^{-5}$ 
\cite{Bennett:gg,Spergel:2003cb,Netterfield:2001yq}. 
Encoded in the statistical correlations of these tiny temperature anisotropies 
are the statistical features of a primordial seed of very small cosmological 
density anisotropies in the, otherwise, homogeneous and isotropic universe at 
the instant of decoupling. Current cosmological models work within the paradigm that 
large scale mass structures in the distribution of matter in the present universe are 
the result of gravitational evolution operating since the time of matter radiation
equality on that primordial seed (for a recent claim of observational evidence
of this correlation see \cite{Eisenstein:2005su}). 

The most characteristic statistical feature of the primordial seed of 
density anisotropies at decoupling is its scale invariance over
cosmologically large comoving scales \cite{HZ}: 
the variance $(\Delta M_V)^2$ of density anisotropies over volumes $V$ of 
cosmological comoving size $L$ is proportional to the area 
$S \sim L^2$ of the surface that bounds the considered region,

\begin{equation}
\label{scaleinvariance}
(\Delta M_V)^2 \simeq \kappa^2 \rho^2_{eq}\ H^{-4}_{eq} S.
\end{equation}
The dimensionless numerical factor $\kappa \sim 10^{-5}$ measures the global
amplitude of the primordial density anisotropies in terms of the cosmological 
density at decoupling $\rho_{eq} = \frac{3 H^2_{eq}}{8\pi G}$ and the comoving 
causal horizon $H^{-1}_{eq}$ at that time. Comoving scales of 
cosmological size are much larger than this horizon, 

\begin{equation}
\label{cosmologicalscales}
H^{-1}_{eq}\ \ll\ L\ \simlt\ H^{-1}_0. 
\end{equation}
The upper bound $H^{-1}_0 \simeq 10^4 H^{-1}_{eq}$ is the comoving size of the 
present causal horizon. For the sake of simplicity we have chosen the scale factor 
of the expansion of the universe at decoupling $a(t_{eq}) = 1$ as reference. 

Scale invariant macroscopic density anisotropies (\ref{scaleinvariance}) 
produce a gravitational potential independent of the scale $L$,

\begin{equation}
\label{gravpotential}
(\Delta \phi|_L)^2 = G^2 \frac{(\Delta M_V)^2}{L^2} \simeq \kappa^2.
\end{equation}
This feature is observable, through the Wolfe-Sachs effect \cite{WolfSachs}, 
as the renowned Harrison-Zeldovich {\it plateau} in the spectrum of 
temperature anisotropies in the CMBR at angular scales larger than the angle 
subtended on the sphere of last scattering by the comoving horizon at 
decoupling, 

\begin{equation}
\frac{\Delta T(\theta)}{T} \simeq \frac{1}{3} \Delta \phi \simeq \kappa,
\hspace{0.2in} at \hspace{0.2in} \theta \gg 1^o.
\end{equation}
The amplitude of temperature fluctuations over large angular scales is 
roughly constant. 

\subsection {The prevalent analysis.}

The pattern (\ref{scaleinvariance},\ref{cosmologicalscales}) of scale invariant 
primordial density anisotropies at decoupling over comoving volumes of size 
much larger than the causal horizon at that time has puzzled cosmologists since 
long ago. Theoretical analysis reproduced in textbooks as 
well as research papers \cite{Peebles,YaZeldovich,Robinson:1995xr} seem to 
conclusively prove that the pattern cannot be reproduced in the (otherwise very 
successful) standard FRW cosmology without violating causal correlations over 
scales  beyond the horizon. The argument goes as follows (see, for example, Eq. 
9.1, 9.2 and 9.3 in \cite{Mukhanov:1990me}). 

The variance of macroscopic density anisotropies at decoupling over 
volumes $V$ of comoving size $L$, which is related to their 
power spectrum ${\cal P}(k)$ by the equality (see section II.B) 

\begin{equation}
\label{variance}
(\Delta M_V)^2 = \rho_{eq}^2 \int \frac{d^3{\vec k}}{(2\pi)^3}\ {\cal P}(k) \left|
\int_V d^3{\vec x}\ e^{-i{\vec k}\cdot{\vec x}}\right|^2,
\end{equation}
is approximated by the expression:

\begin{equation}
\label{approximation}
(\Delta M_V)^2 \simeq \rho_{eq}^2 {\cal P}(k \sim 1/L)\ V,
\end{equation}
after noticing that

\begin{eqnarray}
\nonumber
\left|\int_V d^3{\vec x}\ e^{-i{\vec k}\cdot{\vec x}}\right|^2 & \simeq & V^2
\hspace{0.3in} for \hspace{0.3in} k \simlt 1/L, \\
\label{GEOfactor}
\left|\int_V d^3{\vec x}\ e^{-i{\vec k}\cdot{\vec x}}\right|^2 & \ll & V^2
\hspace{0.3in} for \hspace{0.3in} k \gg 1/L.
\end{eqnarray}
Approximation (\ref{approximation}) relies on the suppression of the
geometric factor (\ref{GEOfactor}) for modes $k \gg 1/L$ to estimate that the
contribution of these modes to macroscopic inhomogeneities (\ref{variance}) over comoving volumes 
of size $L$ is negligible and conclude that the dominant contribution comes 
necessarily from density fluctuations in modes $k \simlt 1/L$ with comparable 
comoving wavelength. 
 
According to expression (\ref{approximation}) the pattern 
(\ref{scaleinvariance},\ref{cosmologicalscales}) of scale invariant primordial 
density anisotropies over cosmologically large comoving volumes 
corresponds univocally to a power spectrum at decoupling

\begin{equation}
\label{linearpower}
{\cal P}_*(k) \simeq (\kappa^2 H^{-4}_{eq})\ k, \hspace{0.3in} for 
\hspace{0.3in} k \ll H_{eq},
\end{equation}
which decreases linearly over the range of modes with cosmologically short 
comoving momenta. To check this inference plug
(\ref{linearpower}) into (\ref{approximation}) to reproduce the scale
invariant pattern (\ref{scaleinvariance},\ref{cosmologicalscales}):
$(\Delta M_V)_*^2 \simeq \rho^2_{eq} {\cal P}_*(k \sim 1/L)\ V \simeq 
\kappa^2\ \rho_{eq}^2 H^{-4}_{eq} \frac{1}{L}\ V \sim 
\kappa^2\ \rho_{eq}^2\ H^{-4}_{eq}\ S$.

Inference (\ref{linearpower}) is bewildering within the theoretical
framework of standard FRW cosmology. The power spectrum of density
anisotropies that standard physics operating during 
the radiation dominated expansion of the universe can produce at decoupling is 
known to be suppressed much strongly than linearly (\ref{linearpower}) over the 
range $k \ll H_{eq}$. The suppression is known to be at least 
quartic \cite{Robinson:1995xr},
 
\begin{equation}
\label{causalpower}
{\cal P}(k) \le {\cal P}(k \sim H_{eq}) \left(\frac{k^4}{H_{eq}^4}\right), 
\hspace{0.3in} for \hspace{0.3in} k \ll H_{eq}.
\end{equation}
This feature is forced by the constrain of global mass (energy) conservation
and by the causal structure of the FRW universe, in which sites separated by 
cosmologically large comoving distances have not ever been in causal contact 
before decoupling.  
Hence, (\ref{causalpower}) applies to the power spectrum of density anisotropies 
generated by any mechanism operating in the cosmic baryon-photon plasma 
before decoupling as long as it does not violate causality. 

A direct evaluation, according to approximation (\ref{approximation}), of the 
variance of cosmological density anisotropies at decoupling that correspond 
to the power spectrum (\ref{causalpower}) leads to 

\begin{equation}
\label{causalflaw}
(\Delta M_V)^2 \simeq (\Delta M_{V_{eq}})^2 \frac{H^{-1}_{eq}}{L}.
\end{equation}
According to this estimation, the variance of density anisotropies that  
standard physics produce at decoupling over cosmologically large comoving volumes 
$V$ of size $L \gg H^{-1}_{eq}$ is predicted to be suppressed by a factor 
$H^{-1}_{eq}/L$ with respect to the variance of the anisotropies 
$(\Delta M_{V_{eq}})^2$ that the same physics generate at that time 
over comoving volumes $V_{eq}$ of the size of the 
causal horizon. The manifest disagreement between the theoretical 
prediction (\ref{causalflaw}) and the observed scale invariant pattern 
(\ref{scaleinvariance}) is known in the literature as the {\it origin of 
structures problem} of standard FRW cosmology. 

In order to solve this problem cosmologists have worked out during the last 
twenty years a quite unexpected cosmological paradigm: a period of exponential 
inflation when the temperature of the universe was close to the Planck scale
would have supposedly preceded the standard radiation dominated expansion of the
universe \cite{Turner:2002ts}. It is claimed that a very early period of 
cosmological inflation would naturally explain the exceptional conditions of the universe at 
the beginning of the standard FRW expansion: namely, its huge 
size and its flatness, as well as its homogeneity and isotropy over 
cosmologically large comoving scales. In addition, the density anisotropies 
present in the very first instants of the history of the universe would be stretched 
by the inflation to comoving cosmological scales. At decoupling these
anisotropies would have a characteristic linear power spectrum 

\begin{equation}
\label{powerinflation}
{\cal P}_{inflat}(k) \sim {\cal A}\ k, \hspace{0.3in} for \hspace{0.3in} k \ll H_{eq}
\end{equation}
over the range of cosmologically short comoving momenta.
According to inference (\ref{linearpower}) these anisotropies would be scale 
invariant, $(\Delta M_V)_{inflat}^2 \simeq \rho^2_{eq} {\cal P}_{inflat}(k \sim 1/L)\
V \simeq \rho^2_{eq} {\cal A}\ S$. The global amplitude of the anisotropies 
left by inflation is fixed by the parameter
${\cal A}$. In most models of inflation, ${\cal A}$ is 
an adjustable parameter that depends on the scale at which inflation takes 
place. This parameter is tuned to the value 
${\cal A} \simeq \kappa^2\ H^{-4}_{eq}$ by additional constrains on 
the inflationary setup. 
 
The widely accepted conclusion of the analysis that we have briefly reviewed
is that primordial scale invariant cosmological anisotropies 
(\ref{scaleinvariance},\ref{cosmologicalscales}) are
a pristine signature of inflation, which cannot have been altered 
by anisotropies (\ref{causalflaw}) generated by standard physics during the 
radiation dominated expansion of the universe. 
Inflation is currently regarded by most cosmologists as the natural solution to 
the {\it origin of structures problem} of FRW cosmology, even though multiple technical problems
(mostly related with the need to causally connect two temperature scales 
$T_{eq} \sim 1$~eV and $T_{Planck} \sim 10^{18}$~GeV separated by 28 orders 
of magnitude) have prevented the inflationary paradigm to get successfully 
implemented in models of particle cosmology. 

\subsection{An unjustified approximation.}

The motivation of this paper is to contend that the above analysis of the 
{\it origin of structures problem} of standard FRW cosmology is flawed by an 
unjustified approximation. We point out that the estimation (\ref{causalflaw}) 
of the variance of cosmological density anisotropies produced at decoupling by 
standard physics (\ref{causalpower}) is deceptive and leads to erroneous 
conclusions. The estimation (\ref{causalflaw}) is based on approximation 
(\ref{approximation}), which
relies on the suppression of the geometric factor (\ref{GEOfactor}) in modes 
$k \gg 1/L$ in order to neglect the contribution of these modes to the 
integral expression (\ref{variance}) for the variance of density anisotropies 
over comoving volumes of size $L$. According to estimation (\ref{causalflaw}), 
the suppression of the geometric factor in modes $k \gg 1/L$ implies that the largest
contribution to the integral expression (\ref{variance}) must come from modes $k \simlt 1/L$.
We notice that this argument is invalid when the considered comoving volume of 
integration is cosmologically large $L \gg H^{-1}_{eq}$, because the power 
spectrum (\ref{causalpower}) of density anisotropies left at decoupling by 
standard physics is greatly enhanced in modes $k \simeq H_{eq} \gg 1/L$ and 
suppressed in modes $k \simlt 1/L$,    

\begin{equation}
\label{powersuppression}
{\cal P}(k \sim 1/L) \simlt 
{\cal P}(k \sim H_{eq})\
\left(\frac{H^{-1}_{eq}}{L}\right)^4, \hspace{0.3in} for \hspace{0.3in} L \gg
H^{-1}_{eq}.
\end{equation} 
In order to get a reliable estimation of the contribution of modes 
$k \simgt H_{eq}$ to the variance (\ref{variance}) of density 
anisotropies produced at decoupling by standard physics (\ref{causalpower}) 
over cosmologically large comoving volumes of size $L \gg H^{-1}_{eq}$, 
we must compare the suppression of the geometric factor (\ref{GEOfactor}) in 
these modes against their enhanced power spectrum (\ref{powersuppression}).  

Indeed we have proven in \cite{oaknin1}, Section III, that approximation 
(\ref{approximation}) does not give a correct estimation of the variance 
(\ref{variance}) of cosmological density anisotropies 
whenever the power spectrum ${\cal P}(k) \sim k^n$ 
decreases faster than linearly, $n \ge 1$, over the range $k \ll H_{eq}$
of comoving modes with cosmologically large comoving wavelength. 
For these power spectra the largest contribution to the variance of 
cosmological density inhomogeneities comes from modes $k \simgt H_{eq}$ with 
comoving wavelength within the horizon, while the contribution from modes 
$k \simlt 1/L$ with cosmologically large comoving wavelength is negligible. 
Approximation (\ref{approximation}) carelessly neglects the contribution 
of modes with comoving wavelength within the horizon to the variance of 
density anisotropies over cosmologically large comoving volumes.
This forgotten contribution accounts for random density fluctuations at the 
boundary surface of the cosmological volume of integration.
Once this contribution is taken into account the variance 
(\ref{variance}) of cosmological density anisotropies associated to the 
power spectra (\ref{causalpower}) is shown to be scale invariant, 

\begin{equation}
\label{scaling}
(\Delta M_V)^2 \simeq (\Delta M_{V_{eq}})^2 
\left(\frac{L}{H^{-1}_{eq}}\right)^2, \hspace{0.3in} for \hspace{0.3in}
L \gg H^{-1}_{eq}.
\end{equation}
This theoretical prediction mends the flawed estimation (\ref{causalflaw}).
In next Section II, subsections II.B and II.C, we prove and discuss this 
prediction in clear physical terms.

In summary, we claim that standard physics operating during the radiation dominated 
expansion of the universe generate at decoupling scale invariant density 
anisotropies over cosmologically large comoving volumes. The cosmological
anisotropies are produced by random density fluctuations with comoving 
wavelength shorter than the comoving causal horizon at the time.
According to prediction (\ref{scaling}), the pattern 
(\ref{scaleinvariance},\ref{cosmologicalscales}) of scale invariant primordial 
cosmological anisotropies is not bewildering in the framework of standard FRW
cosmology. On the contrary, this pattern is the clear signature of the strong 
suppression of the power spectrum of standard physics (\ref{causalpower}) in 
modes with cosmologically large comoving wavelength. Scale invariance 
is dictated by the causality constrains of standard FRW cosmology. 
This conclusion obviously contradicts the current wisdom in cosmology, 
expressed in the flawed Eq. (\ref{causalflaw}). 

Equation (\ref{scaling}) also proves that the inference (\ref{linearpower}) of 
a linear power spectrum of primordial cosmological density anisotropies at the 
time of decoupling is absolutely unjustified. Scale invariance 
(\ref{scaleinvariance},\ref{cosmologicalscales}) of primordial cosmological
density anisotropies at decopling only implies that their power spectrum
decreases faster than linearly over the range of cosmologically short comoving 
momenta. The power spectrum (\ref{causalpower}) fulfills this constrain.

\subsection{A revised analysis.}

The global amplitude of the scale invariant cosmological density anisotropies 
(\ref{scaling}) produced at decoupling by standard physics operating during the 
radiation dominated expansion of the universe is fixed by their power spectrum 
${\cal P}(k)$ in modes $k \simgt H_{eq}$ with comoving wavelength shorter than 
the causal horizon at that time. Hence, their amplitude is 
proportional to the amplitude $(\Delta M_{V_{eq}})^2 \simeq \rho^2_{eq}
{\cal P}(k \sim H_{eq})\ V_{eq}$ of the inhomogeneities that the same physics 
produce over comoving volumes $V_{eq}$ of the size of this causal 
horizon. A comparison of (\ref{scaling}) with (\ref{scaleinvariance}) implies 
that it is necessary

\begin{equation}
\label{Peebles}
(\Delta M_{V_{eq}}) \simeq \kappa\ \rho_{eq} V_{eq}
\simeq 10^{-5} \rho_{eq} V_{eq},
\end{equation}
for standard physics operating during the radiation dominated 
expansion of the universe to produce at decoupling scale invariant 
cosmological density anisotropies with global amplitude comparable to 
that of the primordial scale invariant density inhomogeneities 
imprinted in the CMBR. This conclusion reveals an unexplored, but feasible, 
scenario for the origin of primordial cosmological structures within the 
framework of FRW cosmology \cite{oaknin3}. 
This scenario had been previously disregarded in the literature, in favour of 
the inflationary paradigm, due to the flawed estimation (\ref{causalflaw}).

During the radiation dominated expansion of the universe the cosmic plasma
is a dense and opaque soup of coupled baryons and photons in thermal 
equilibrium. Density (pressure) waves can propagate in this fluid with 
sound velocity $c_s \simeq \frac{1}{\sqrt{3}}$ (in natural units). These waves 
homogenize the cosmic density field, eroding all random density fluctuations 
with comoving wavelength much shorter than the sound causal horizon at the time. 
Obviously, causality constrains prevents this mechanism to operate over 
comoving distances larger than the sound causal horizon. Hence, the power 
spectrum of the density anisotropies left by standard physics 
in the cosmic plasma at decoupling is peaked at $k \simeq H_{eq}$ and suppressed 
${\cal P}(k) \simlt {\cal P}(k \sim H_{eq})\ \left(\frac{H_{eq}}{k}\right)^3$ 
over modes $k \gg H_{eq}$ by the action of these non-linear wave phenomena. 
We have discussed above (\ref{causalpower}) that the power spectrum 
of the density anisotropies left by standard physics is also 
suppressed over modes $k \ll H_{eq}$ because of causal constrains. Such power 
spectrum corresponds to a pattern of macroscopic density anisotropies 
$(\Delta M_V)^2 \simeq (\Delta M_{V_{eq}})^2 \left(\frac{V}{V_{eq}}\right)^2$ 
over volumes of comoving size $L \simlt H^{-1}_{eq}$, which turns scale invariant
(\ref{scaling}) over volumes of comoving size $L \gg H^{-1}_{eq}$.
We will show in next Section II, subsection II.E, that 
the amplitude (\ref{Peebles}) is roughly given by

\begin{equation}
\label{ampl}
(\Delta M_{V_{eq}})^2 \simeq 2\langle [\rho({\vec x})]^2 - 
\rho^2_{eq}\rangle\ V^2_{eq} = 2\rho^2_{eq} \left(\int \frac{d^3{\vec k}}{(2\pi)^3}
{\cal P}(k)\right)\ V^2_{eq}, 
\end{equation}  
where $\rho({\vec x})$ is the cosmological density field at decoupling.
The dimensionless factor can be estimated by 
$\int \frac{d^3{\vec k}}{(2\pi)^3} {\cal P}(k) \simeq 
{\cal P}(k \sim H_{eq})\ H^3_{eq}$. Thus, Eq. (\ref{Peebles}) is satisfied if 

\begin{equation}
\label{amp2}
{\cal P}(k \sim H_{eq})\ H^3_{eq} \simeq \kappa^2 \simeq 10^{-10}.
\end{equation}

The dimensionless factor $\int \frac{d^3{\vec k}}{(2\pi)^3} {\cal P}(k)$ 
that fixes (\ref{ampl}) the global amplitude of the cosmological density 
anisotropies at decoupling (\ref{scaling}) ia a measure also of the entropy of 
the fluctuations. If the radiation dominated expansion of the universe is 
adiabatic and the entropy of the density fluctuations is conserved, then the 
global amplitude of the cosmological density anisotropies at decoupling is set 
by the dynamics of the universe at a much earlier stage.
We discuss condition (\ref{ampl}) in further detail below in Section II, 
subsection II.E.

The implications of this revised analysis are far reaching.
The standard FRW cosmological framework that we are discussing here assumes, 
implicitly, that the exceptional initial conditions at the beginning of the 
radiation dominated expansion of the universe (namely, its huge size and flatness, 
as well as its homogeneity and isotropy over cosmologically large comoving 
scales) have been set by some unspecified mechanism ({\it e.g.} inflation) at 
some earlier stage. According to the prevalent analysis of the origin
of primordial cosmolgical structures, see subsection I.A, the mechanism that fixes the initial
homogeneous and isotropic conditions of the universe needs to be tuned to seed 
at once density fluctuations with cosmologically large comoving wavelength
and the right amplitude, see Eq. (\ref{powerinflation}). The difficulties to get 
the mechanism appropriately tuned have an obvious origin: cosmological homogeneity and 
isotropy means, by definition, that all density fluctuations with cosmologically 
large comoving wavelength should be negligibly small. 
We realize now that, according to Eq. (\ref{scaling},\ref{Peebles}), 

\begin{itemize}
\item The primordial cosmological density anisotropies 
(\ref{scaleinvariance},\ref{cosmologicalscales}) have not to be
necessarily seeded by the same mechanism that fixes the initial homogeneous 
and isotropic conditions at the beginning of the standard FRW expansion
of the universe. The anisotropies may be seeded later on, during the radiation 
dominated expansion of the universe. Thus, we should consider as feasible models of 
inflation (or any other mechanism) that can fix the initial size, 
flatness, homogeneity and isotropy of the FRW universe, and leaves 
negligibly small density fluctuations with cosmologically large wavelength.
Such models should be easier to implement in models of particle cosmology.  
The price to pay is obvious: the physics of such a mechanism 
would not be readily testable on the spectrum of temperature anisotropies in the 
CMBR. 

\item Scale invariant primordial density anisotropies 
(\ref{scaleinvariance},\ref{cosmologicalscales}) 
over cosmologically large comoving volumes are not necessarily a pristine signature of relic 
inhomogeneities left over by the mechanism that fixed the initial conditions at the beginning of the
radiation dominated expansion of the universe (e.g. inflation). Even in the case 
that relic scale invariant cosmological inhomogeneities were left by that mechanism at the beginning of the radiation
dominated expansion, standard physics operating later 
on produce scale invariant cosmological anisotropies that are overlaid upon 
them by the time of decoupling.
\end{itemize}


\section{Scale invariant density anisotropies.}

The main aim of this paper is to prove that the characteristic scale 
invariance of the primordial cosmological density anisotropies at 
decoupling (\ref{scaleinvariance},\ref{cosmologicalscales}) is an unequivocal  
signature of the standard FRW cosmology. In the standard cosmology the causal
comoving horizon grows monotonically with cosmic time. Comoving scales
of cosmological size are much larger than the comoving causal horizon at 
decoupling. The short comoving causal horizon prevents the generation of 
density fluctuations with cosmologically large comoving wavelength during the 
radiation dominated expansion of the universe (\ref{causalpower}). 
As a consequence the dominant contribution to density 
anisotropies at decoupling over cosmologically large comoving volumes comes 
from random density fluctuations with comoving wavelength shorter than the 
horizon. This contribution corresponds to random density fluctuations at the 
boundary surface of the volume of integration and, hence, the variance of these
anisotropies grows with the area of the boundary surface. Below we compare 
the geometric dependence of the variance of the primordial cosmological 
density anisotropies on the boundary area (\ref{scaleinvariance}) with the dependence 
of the variance of thermal anisotropies in systems in equilibrium on some power 
$\beta \in [1, 2)$ of the volume of integration, $(\Delta M_V)^2 \sim V^{\beta}$.

It has been formally proved \cite{Beck} that in any homogeneous and 
isotropic statistical system the variance $(\Delta M_V)^2$ of the spatial 
anisotropies of any extensive magnitude $M_V$ over macroscopically large, but finite, 
volumes $V$ must grow with their size $L$ as $(\Delta M_V)^2 \sim L^{\alpha}$, 
with $\alpha \ge 2$. The derivation of equation (\ref{twocontributions}) below 
is a brief proof of this result. Scale invariant statistical systems like 
(\ref{scaleinvariance}), for which $\alpha =2$, have been dubbed 
for this reason {\it superhomogeneous} \cite{Gabrielli:2001xw}. We can say that, 
in this sense, scale invariant systems are the most ordered homogeneous and 
isotropic statistical systems and thermal systems in equilibrium, for which 
$\alpha \ge 3$, are more disordered than scale invariant systems.

Scale invariant anisotropies are known in mathematics, for example, 
in the distribution over the complex plane of zeroes of statistical analytic functions
\cite{sodin}. In physics, scale invariant anisotropies happen generically in homogeneous
and isotropic statistical systems with very long relaxation times to thermal 
equilibrium. The reader familiar with the arguments that justify Eq. 
(\ref{causalpower}) will appreciate the following illustrative example.

\subsection{A simple example of scale invariant anisotropies.} 

Let consider an homogeneous and isotropic distribution of a large 
number of individuals over a very large (infinite) domain 
$\Omega \subset {\bf R}^3$ and assume that this initial distribution of 
individuals does not contain random density fluctuations with wavelength longer 
than the shortest accessible resolution scale $d$.

We now let each individual to walk randomly and assume the same 
statistical features for the individual random walks everywhere in $\Omega$. 
The {\it typical} distance $D(t) \equiv \langle |{\vec r}(t) - {\vec r}_0|^2 
\rangle^{1/2}$ between the actual position of each individual at time $t$ 
with respect to its original position grows monotonically with the time interval 
$D(t) = \sqrt{\gamma}\ t^{1/2}$ as a function determined by the specific properties 
of the stochastic process. Let us call the distance $D(t)$ a particle horizon: 
at any instant each individual is constrained to be located 
within a finite region of typical size $D(t)$ around its original position. 
Let us also assume that we have waited enough, such that $D(t) \gg d$. 

Let us denote by $N_V$ the number of individuals contained within a 
macroscopic volume $V \subset \Omega$ at time $t$. Macroscopic volumes are 
defined as volumes of typical size $L \simgt d$. Random fluctuations in 
the extensive statistical magnitude $N_V$ result from the individual 
random walks. We wish to characterize their variance $(\Delta N_V)^2$
as a function of the size $L$ of the considered macroscopic volume. 

If the macroscopic volume has a typical size $L$ in the range $d \simlt L 
\simlt D(t)$ (of the order of the particle horizon $D(t)$ or shorter) the 
fluctuations are obviously thermal $(\Delta N_V)^2 \propto V$, as we have 
nothing but an example of brownian motion of independent particles in a 
macrocanonical ensemble. Indeed the particle horizon $D(t)$ is not evident 
locally, over volumes of smaller size. The {\it temperature} of these thermal 
anisotropies is dictated by the statistical features of the random walks. If
the laws of the individual random walks are invariant under spatial translations 
and rotations in $\Omega$ the system remains statistically homogeneous and
isotropic at any time. The {\it temperature} of the local anisotropies may
change with time, but at any instant it is the same over the whole space 
$\Omega$ (even though the existence of a finite
particle horizon implies that the statistical system cannot have relaxed to 
thermal equilibrium over scales beyond it).

The existence of a finite particle horizon is manifest in the fluctuations in 
the number of individuals $N_V$ in volumes $V$ of size $L \gg D(t)$.
The density anisotropies over volumes much larger than the particle horizon
are scale invariant $(\Delta N_V)^2 \sim S$, similar to the primordial 
cosmological density anisotropies at decoupling (\ref{scaleinvariance}) that 
are the focus of our interest in this paper. In order to justify this claim let 
us first define the cover of width $D(t)$ of the boundary surface of a macroscopic 
finite volume $V$, and denote it as $\partial_{D(t)} V$, as the union of all balls of radius
$D(t)$ around any point $x \in \partial V$: the cover $\partial_{D(t)} V$ of a
volume $V$ is the set of all points whose distance to its boundary surface 
$\partial V$ is smaller than $D(t)$.
Individuals that are originally located out of this cover cannot contribute 
at the instant $t$ to fluctuations in the number of individuals within the 
volume $V$, even though they also are walking randomly. 
Only individuals that were originally located within $\partial_{D(t)} V$
can produce, when they cross the border $\partial V$, fluctuations
in the total number of individuals in $V$. That is, only the degrees of 
freedom in $\partial_{D(t)} V$ are apparent in these fluctuations. 
Naturally, then, the anisotropies that the random walks of these degrees of 
freedom induce in the number of particles $N_V$ are scale invariant 
$(\Delta N_V)^2 \sim D(t)\ S \sim D(t)\ L^2 $, rather than thermal 
$(\Delta N_V)^2 \sim V \sim L^3$.

The power spectrum of the number density anisotropies generated by the
random walks of individual particles has been described in the literature 
\cite{YaZeldovich,Robinson:1995xr},

\begin{eqnarray}
\label{powerbrownian}
{\cal P}(k; t) = \left\{
\begin{array}{cc}
C, & \hspace{0.3in} for \hspace{0.2in} k \simgt D(t)^{-1} \\ 
C \left[k\ D(t)\right]^4, & \hspace{0.3in} for \hspace{0.2in} 
k \ll D(t)^{-1},
\end{array}
\right.
\end{eqnarray}
where $C$ is a constant determined by the specific features of the random walks.
This power spectrum is roughly constant over the range of modes with
wavelength shorter than the particle horizon $D(t)$, but it decreases
as $\sim k^4$ for modes with longer wavelength. The suppression of this power 
spectrum in modes with wavelength longer than the horizon $D(t)$ is similar
to the suppression of the power spectrum (\ref{causalpower}) of cosmological
density anisotropies left at decoupling by standard physics operating during 
the radiation dominated expansion of the universe.
Hence, on scales $L \gg D(t)$ condition (\ref{powersuppression}) holds for power 
spectrum (\ref{powerbrownian}), with $H^{-1}_{eq}$ replaced by $D(t)$. 
In fact, this condition defines the presence of a causal horizon in 
homogeneous and isotropic statistical systems and implies that the system has 
not reached thermal equilibrium over scales beyond it. 

It is clear from this example that the scale invariance of spatial 
random anisotropies over volumes of size 
$L \gg D(t)$ is a direct consequence of the fact that individual degrees of 
freedom are constrained within a finite particle horizon $D(t)$. The horizon 
prevents the development of random density fluctuations with longer wavelength.
This constrain is expressed through the suppression of the power spectrum 
(\ref{powerbrownian}) in modes $k \ll D(t)^{-1}$.
As a result of the strong suppression of the power 
spectrum in these modes, the largest contribution to the integral expression
(\ref{variance}) for the variance of density anisotropies over 
volumes of size $L \gg D(t)$ comes from modes $k \simgt D(t)^{-1}$. Their 
contribution is scale invariant because it corresponds to density fluctuations 
at the boundary surface of the considered volume of integration.

We wish to remark that the contribution from modes $k \simgt D(t)^{-1}$ to 
macroscopic density anisotropies over volumes of size $L \gg D(t)$
is not a {\it mathematical pathology} associated to a boundary surface that 
needs to be resolved with infinite precision, as sometimes claimed in the 
literature in cosmology. Scale invariant anisotropies are produced by boundary 
fluctuations with typical wavelength $d \ll \lambda \simlt D(t)$, much shorter 
than the size $L$ of the considered volume of integration but still much longer 
than the resolution scale $d$ of the boundary surface.

\subsection{The formal arguments.}
   
We now wish to formalize these arguments using simple tools and
consider an statistical density field on ${\bf R}^3$, 

\begin{equation}
\label{density}
\rho({\vec x}; d) = \rho_{eq} + \rho_{eq} \int \frac{d^3{\vec k}}
{(2\pi)^3}\ \delta_{\vec k}(d)\ e^{-i{\vec k}\cdot{\vec x}},
\end{equation}
to describe, in comoving coordinates, the cosmological density field at the 
instant of decoupling resolved at a certain finite length scale $d$. 
The scale factor of the universe at decoupling $a(t_{eq}) = 1$ is chosen
as reference. The density field is assumed to be homogeneous and isotropic, 

\begin{equation}
\label{uniformity}
\langle \rho({\vec x}; d) \rangle = \rho_{eq}
\end{equation}
(or equivalently, $\langle \delta_{\vec k}(d) \rangle = 0$), up to tiny 
statistical fluctuations. We assume that the random modes 
$\delta^*_{\vec k}(d) = \delta_{-{\vec k}}(d)$ are independent
statistical complex variables with gaussian distribution,

\begin{equation} 
\langle \delta^*_{{\vec k}_1}(d) \delta_{{\vec k}_2}(d) 
\rangle = (2\pi)^3\ {\cal P}({\vec k_1}; d)\ \delta^3({\vec k}_1 - {\vec k}_2).
\end{equation}
The function ${\cal P}({\vec k}; d)$ is known as the power spectrum of the statistical 
fluctuations and is always positive ${\cal P}({\vec k};d) \ge 0$. In isotropic
statistical systems the power spectrum is a function 
${\cal P}({\vec k};d) = {\cal P}(k; d)$ only of the modulus
of the wave vector. 

The finite resolution scale of the density field is introduced in the
description through a convolution kernel (window function) with typical width 
$d > 0$, $\rho({\vec x}; d) = \int d^3{\vec y}\ W({\vec x} - {\vec y}; d)\
\rho({\vec y})$ (see section VIII in \cite{oaknin1}). The specific features of
the window function $W({\vec x} - {\vec y}; d)$ are not of particular interest
here, other than it introduces a natural ultraviolet cutoff in the
the power spectrum ${\cal P}(k; d) = {\cal P}(k) 
|{\widetilde W}(k; d)|^2$ over scales shorter than the
resolution length: $|{\widetilde W}(k; d)|^2 \ll 1$, for $k \gg 
d^{-1}$ and $|{\widetilde W}(k; d)|^2 \sim 1$, for $k \simlt 
d^{-1}$. In other words, the convolution kernel in the definition of
$\rho({\vec x};d)$ simply integrates out all modes with wavelength shorter
than the resolution scale. 

Macroscopic volumes are defined as volumes whose typical size is much larger 
than the resolution scale $d$. According to the central limit theorem the 
statistical extensive magnitudes 

\begin{equation}
M_V(d) = \int_V d^3{\vec x}\ \rho({\vec x}; d)
\end{equation} 
have gaussian distribution. Hence, they are
completely specified by their first two momenta: the average value 
$\langle M_V \rangle = \rho_{eq}\ V$ and the variance of its random fluctuations 
$(\Delta M_V)^2$, which is related to their power spectrum by expression 
(\ref{variance}).

Expression (\ref{variance}) emphasizes that the variance of random density
anisotropies over macroscopic volumes of any size $L$ receives positive 
contributions from random fluctuations in all modes $\delta_{\vec k}(d)$. 
We want to evaluate separately the contribution from modes $k \gg L^{-1}$
and compare it to the contribution from modes $k \simlt L^{-1}$,

\begin{equation}
\label{spectrum}
(\Delta M_V)^2 \simeq \rho_{eq}^2 \int_0^{k \simlt L^{-1}} 
\frac{d^3{\vec k}}{(2\pi)^3} {\cal P}(k) \left|\int_V d^3{\vec x}\ 
e^{-i{\vec k}\cdot {\vec x}}\right|^2 + 
\rho_{eq}^2 \int_{k \gg L^{-1}}^{k \simlt d^{-1}} 
\frac{d^3{\vec k}}{(2\pi)^3} {\cal P}(k) \left|\int_V d^3{\vec x}\ 
e^{-i{\vec k}\cdot {\vec x}}\right|^2.
\end{equation}
The first contribution, from modes $k \simlt 1/L$, coincides with
approximation (\ref{approximation}): 

\begin{equation}
\label{bulk}
\rho_{eq}^2 \int_0^{k \simlt L^{-1}} \frac{d^3{\vec k}}{(2\pi)^3} {\cal P}(k) 
\left|\int_V d^3{\vec x}\ e^{-i{\vec k}\cdot {\vec x}}\right|^2 \simeq 
\rho_{eq}^2 \frac{1}{V} {\cal P}(k \sim 1/L)\ V^2 \sim 
\rho_{eq}^2 {\cal P}(k \sim 1/L)\ V.
\end{equation}
The second contribution, from modes $k \gg L^{-1}$, is neglected in
approximation (\ref{approximation}). This contribution is the focus of our 
interest in this paper because it is scale invariant,

\begin{equation}
\label{boundary}
\rho_{eq}^2 \int_{k \gg L^{-1}}^{k \simlt d^{-1}} 
\frac{d^3{\vec k}}{(2\pi)^3} {\cal P}(k) \left|\int_V d^3{\vec x}\ 
e^{-i{\vec k}\cdot {\vec x}}\right|^2 \simeq
\frac{\rho_{eq}^2}{4\pi^2} \left[\int_{k \gg L^{-1}}^{k \simlt d^{-1}} 
d^3{\vec k}\ \frac{{\cal P}(k)}{k^4}\right]\ S.
\end{equation}
As above, $S$ stands for the area of the surface that bounds the volume 
of integration $V$. 
Equation (\ref{boundary}) is very simple to derive in 
the case when the volume of integration $V$ is a sphere of radius $L$,

\begin{equation}
\left|\int_V d^3{\vec x}\ e^{-i{\vec k}\cdot {\vec x}}\right|^2 = 
4\pi\ \frac{S}{k^4} \left(\frac{1}{k L} sin(k L) - \hspace{0.03in} cos(k L) 
\right)^2,
\end{equation}
and the power spectrum ${\cal P}(k)$ in the range $k \gg L^{-1}$ is smooth 
over scales $dk \simlt L^{-1}$.
The result can be immediately extended to any other connected volume, assuming 
that the surface that bounds it is smooth and its area does not depend on the 
resolution scale.

In summary, the variance of spatial random anisotropies of extensive 
magnitudes over macroscopic volumes $V$ of any size 

\begin{equation}
\label{twocontributions}
(\Delta M_V)^2 \simeq \rho^2_{eq} {\cal P}(k \sim 1/L)\ V + 
\frac{\rho^2_{eq}}{\pi} \left[\int_{k \gg L^{-1}}^{k \simlt d^{-1}} 
dk\ \frac{{\cal P}(k)}{k^2}\right]\ S, 
\end{equation} 
receives always a bulk contribution from fluctuations with wavelength comparable 
to the size of the volume of integration and an additional scale invariant 
contribution from fluctuations at its boundary surface with much shorter wavelength. 
We notice that the global amplitudes of each of these contributions depend,
respectively, on the power spectrum in modes $k \simlt 1/L$ and $k \gg 1/L$. Obviously,
if the first contribution is negligible compared to the second, then the anisotropies 
are scale invariant.

Let us now consider, in particular, the density anisotropies produced 
at decoupling by standard physics operating during the radiation dominated
expansion of the universe. Their power spectrum (\ref{causalpower}) is strongly 
suppressed in modes $k \ll H_{eq}$ with comoving wavelength much larger than 
the causal horizon. Over cosmologically large comoving volumes, of size 
$L \gg H^{-1}_{eq}$, condition (\ref{powersuppression}) holds.
An explicit evaluation of the two contributions in 
(\ref{twocontributions}) gives

\begin{equation}
\label{C1}
\rho^2_{eq}\ {\cal P}(k \sim 1/L)\ V \simlt \rho^2_{eq}\ {\cal P}(k \sim H_{eq})\ 
\left(\frac{H^{-1}_{eq}}{L}\right)^4\ V \sim \rho^2_{eq}\ {\cal P}(k \sim H_{eq})\
H^{-4}_{eq}\ L^{-1},
\end{equation}
for the bulk contribution  and 

\begin{eqnarray}
\label{C2}
\frac{\rho^2_{eq}}{\pi}\ \left[\int_{k \gg L^{-1}}^{k \simlt d^{-1}} 
dk\ \frac{{\cal P}(k)}{k^2}\right] S \simeq
\frac{\rho^2_{eq}}{\pi}\ \left[\int_{k \gg L^{-1}}^{k \simlt H_{eq}} 
dk\ \frac{{\cal P}(k)}{k^2}\right] S 
\simeq \rho^2_{eq}\ {\cal P}(k \sim H_{eq}) H^{-1}_{eq} L^2,
\end{eqnarray}
for the scale invariant contribution. We notice that (\ref{C1}) corresponds
to the estimation (\ref{causalflaw}) of the variance of primordial cosmological 
density anisotropies at decoupling in standard FRW cosmology, which led to the
formulation of the {\it origin of structures problem}. This estimation neglects
the second contribution (\ref{C2}). Nevertheless, a direct comparison clearly
shows that over volumes of comoving size $L \gg H^{-1}_{eq}$ the second, scale 
invariant contribution (\ref{C2}), from modes $k \simgt H_{eq}$, prevails:

\begin{equation}
\label{laststep}
(\Delta M_V)^2 \simeq \rho^2_{eq}\ {\cal P}(k \sim H_{eq}) H^{-1}_{eq} L^2,
\end{equation}
while the first, bulk contribution (\ref{C1}), from modes $k \simlt 1/L$,
is negligible. The flawed estimation (\ref{causalflaw}) keeps the negligible
contribution (\ref{C1}) and neglects the dominant one (\ref{C2}).

The power spectrum ${\cal P}(k \sim H_{eq})$ in modes with comoving wavelength 
comparable to the comoving causal horizon is not suppressed with respect to 
the power in modes with shorter wavelength. Therefore, over comoving volumes 
$V_{eq}$ of the size of the horizon approximation (\ref{approximation}) 
is valid, $(\Delta M_{V_{eq}})^2 \simeq \rho^2_{eq} {\cal P}(k \sim H_{eq}) V_{eq}
\simeq \rho^2_{eq} {\cal P}(k \sim H_{eq})\ H^{-3}_{eq}$.
Equation (\ref{scaling}) is obtained straightforward from (\ref{laststep}). 

Finally, we wish to remark that in the framework we have laid in this section we 
did not even need to worry about the technical difficulties that appear when 
trying to manage the coupled dynamics of density fluctuations with comoving 
wavelength longer than the causal horizon and space-time metric perturbations 
\cite{Mukhanov:1990me}, because the power spectrum in these modes is strongly 
suppressed and their contribution to cosmological anisotropies is sub-leading.

\subsection{Scale invariant gravitational potential.} 

We have claimed in the Introduction to this paper that scale invariant 
primordial density anisotropies over cosmologically large 
comoving volumes (\ref{scaleinvariance},\ref{cosmologicalscales}) produce 
a statistically constant gravitational potential (\ref{gravpotential})
over cosmologically large comoving distances. In this subsection we prove 
that this result is a consequence of the strong suppression 
(\ref{powersuppression}) of the power spectrum of primordial density 
anisotropies in modes with cosmologically large comoving wavelength. 
 
The statistical gravitational potential at the time of decoupling

\begin{equation}
\label{gravpot}
\phi({\vec x}) = \int \frac{d^3{\vec k}}{(2\pi)^3}\ \phi_{\vec k}\ 
e^{-i{\vec k}\cdot {\vec x}},
\end{equation}
is related to the statistical density field at the time
(\ref{density}) by the equation

\begin{equation}
{\vec \nabla}^2 \phi({\vec x}) = 4\pi G\left(\rho({\vec x}) - 
\rho_{eq}\right).
\end{equation} 
In momentum space the same equation can be written

\begin{equation}
\phi_{\vec k} = -4\pi G \rho_{eq}\ \frac{1}{|{\vec k}|^2}\ \delta_{\vec k}.
\end{equation}
The gravitational field (\ref{gravpot}) is real because

\begin{equation}
\phi_{-{\vec k}}^* = \phi_{\vec k}
\end{equation}
and it is zero everywhere on average $\langle \phi({\vec x}) \rangle = 0$,
because
 
\begin{equation}
\langle \phi_{\vec k} \rangle = 0.
\end{equation}
Thus, the average potential difference between any two points is also zero,

\begin{equation}
\langle \phi(\vec x) - \phi(\vec y) \rangle = 0.
\end{equation}
The two points correlation function of the statistical gravitational
potential is related to the power spectrum of density fluctuations by 
the expression,

\begin{eqnarray}
\nonumber
\langle \phi({\vec x})\ \phi({\vec y}) \rangle = 
\int \frac{d^3{\vec k}_1}{(2\pi)^3} \int \frac{d^3{\vec k}_2}{(2\pi)^3}
\langle \phi^*_{{\vec k}_1} \phi_{{\vec k}_2}\rangle
e^{+i{\vec k}_1\cdot {\vec x}} e^{-i{\vec k}_2\cdot {\vec y}} \\
\nonumber
= (4\pi G \rho_{eq})^2 \int \frac{d^3{\vec k}_1}{(2\pi)^3} \int 
\frac{d^3{\vec k}_2}{(2\pi)^3} \frac{\langle \delta^*_{{\vec k}_1} 
\delta_{{\vec k}_2}\rangle}{|{\vec k}_1|^2 |{\vec k}_2|^2}
e^{+i{\vec k}_1\cdot {\vec x}} e^{-i{\vec k}_2\cdot {\vec y}} \\
= (4\pi G \rho_{eq})^2 \int \frac{d^3{\vec k}}{(2\pi)^3} 
\frac{{\cal P}(|{\vec k}|)}{|{\vec k}|^4}
e^{+i{\vec k}\cdot ({\vec x}-{\vec y})}
\end{eqnarray}
and the variance of random fluctuations in the potential difference between 
any two points is given by,

\begin{eqnarray}
\nonumber
\langle \left[\phi(\vec x) - \phi(\vec y)\right]^2 \rangle = 
2\langle \phi(\vec x)^2 \rangle - \langle \phi(\vec x)\ \phi(\vec y)\rangle
- \langle \phi(\vec y)\ \phi(\vec x)\rangle = \\
\label{gravivariance}
= (4G \rho_{eq})^2 \int dk\ 
\frac{{\cal P}(k)}{k^2} \left(1 - \frac{Sin\left[k|{\vec x}-{\vec y}|\right]}
{k|{\vec x}-{\vec y}|}\right).
\end{eqnarray}
We are interested in the asymptotic behaviour of this variance when 
$|{\vec x}-{\vec y}| \gg H^{-1}_{eq}$, the comoving distance between the
two considered points is cosmologically large. To evaluate the integral 
expression (\ref{gravivariance}) we separate the contribution from modes 
$k \ll H_{eq}$, whose comoving wavelength is much larger than the horizon, 
from the contribution of modes $k \simgt H_{eq}$, whose comoving wavelength 
is within the horizon,

\begin{eqnarray}
\langle \left[\phi(\vec x) - \phi(\vec y)\right]^2 \rangle
\simeq (4G \rho_{eq})^2 \left[\int_0^{k \ll H_{eq}} dk\ 
\frac{{\cal P}(k)}{k^2} \left(1 - \frac{Sin\left[k|{\vec x}-{\vec y}|\right]}
{k|{\vec x}-{\vec y}|}\right) + \int_{k \simgt H_{eq}} dk\ 
\frac{{\cal P}(k)}{k^2}\right]
\end{eqnarray}
The latter contribution is roughly constant $(4G \rho_{eq})^2\ 
\frac{{\cal P}(k \sim H_{eq})}{H_{eq}}$
because $\frac{Sin\left[k|{\vec x}-{\vec y}|\right]} {k|{\vec x}-{\vec y}|} 
\ll 1$, for $k \simgt H_{eq}$ and $|{\vec x}-{\vec y}| \gg H^{-1}_{eq}$. 
We want now to explore the contribution from comoving modes
with cosmologically large wavelength when the power spectrum of density
fluctuations is largely suppressed in this range (\ref{powersuppression}),

\begin{eqnarray}
\nonumber
\int_0^{k \ll H_{eq}} dk\ \frac{{\cal P}(k)}{k^2} \left(1 - 
\frac{Sin\left[k|{\vec x}-{\vec y}|\right]} {k|{\vec x}-{\vec y}|}\right) 
\simlt \frac{{\cal P}(k \sim H_{eq})}{H^4_{eq}} \int_0^{k \ll H_{eq}} dk\ 
k^2 \left(1 - \frac{Sin\left[k|{\vec x}-{\vec y}|\right]} 
{k|{\vec x}-{\vec y}|}\right) \\
\nonumber
\ll \frac{{\cal P}(k \sim H_{eq})}{H_{eq}}.
\end{eqnarray} 
We find again that the contribution from random density fluctuations with
cosmologically large comoving wavelength is suppressed relative to the
contribution from density fluctuations with comoving wavelength wihin the 
horizon. In consequence, $\langle \phi(\vec x) \phi(\vec y) \rangle
\ll \langle \left[\phi(\vec x)\right]^2 \rangle$ when $|{\vec x} - {\vec y}|
\gg H^{-1}_{eq}$ and 

\begin{eqnarray}
\label{statisticalconstant}
\langle \left[\phi(\vec x) - \phi(\vec y)\right]^2 \rangle
\simeq \langle \left[\phi(\vec x)\right]^2 \rangle + \langle
\left[\phi(\vec y)\right]^2 \rangle \simeq
(4G \rho_{eq})^2\ \frac{{\cal P}(k \sim H_{eq})}{H_{eq}}
\end{eqnarray} 
the random gravitational potential is statistically constant over 
cosmologically large comoving distances.

\subsection{Scale invariant anisotropies vs. thermal anisotropies.}

The result (\ref{scaling}) that we have proved in subsection II.B above
is at odds with the intuition 
inferred from the common experience with thermodynamic systems in equilibrium,
in which the variance of random density anisotropies of extensive magnitudes grows 
with the size of the considered region of integration as some power of its 
volume $(\Delta M_V)_{thermal}^2 \sim V^{\beta}$, with $1 \le \beta < 2$. 
In this section we want to compare the bulk dependence of the variance 
of thermal anisotropies with the surface dependence of the variance 
of scale invariant anisotropies. 

Consider a thermodynamic system in equilibrium far from a phase transition. 
Thermal fluctuations in the local densities are described by Poisson two 
points correlation functions $F({\vec x} - {\vec y}) \equiv \frac{1}{\rho^2_0}
\left[\langle \rho({\vec x}) \rho({\vec y}) \rangle - \langle 
\rho({\vec x}) \rangle \langle \rho({\vec y}) \rangle\right] = 
\zeta^2\ d^3\ \delta^3({\vec x} - {\vec y})$, or the more realistic 

\begin{equation}
\label{Poisson}
F(r) = \zeta^2\ \frac{1}{\pi^{3/2}}\ e^{-(r/d)^2}, 
\end{equation}
where $d \simeq 1/T$ is the finite resolution scale and $\zeta$ is a 
dimensionless factor that measures the global amplitude of the fluctuations. 
The corresponding power spectrum,

\begin{equation}
\label{powerthermal}
{\cal P}(k) = \int d^3{\vec r}\ F({\vec r})\ 
e^{+i{\vec k}\cdot {\vec r}} \simeq\ \zeta^2\ d^3
\end{equation}
is roughly constant over the whole range $0 < k \simlt d^{-1}$ of
accessible scales. Hence, over spheres of radius $L$ the variance of spatial 
thermal anisotropies is 

\begin{equation}
\label{varianceT}
(\Delta M_V)^2 = 8\zeta^2\ \rho_0^2\ d^3\ L^3\ 
\int_0^{L/d}\ d(k L)\ \left(\frac{1}{(k L)^2} 
sin(k L) - \frac{1}{k L} cos(k L) \right)^2 \simeq \zeta^2\ \rho_0^2\ d^3\ V.
\end{equation}
It can be readily seen that, independently of the size of the region $L$,
the largest contribution to the last integral comes from modes 
$k L \sim 1$ (the integrand vanishes for $k L \ll 1$ and for $k L \gg 1$) 
and the variance is linearly proportional to the volume of integration, in good 
agreement with the estimation (\ref{approximation}).

The linear dependence of the variance (\ref{varianceT}) 
on the volume of the considered region of integration results from the short 
range of the Poisson two points correlation function (\ref{Poisson}). Poisson 
correlation is characteristic of thermodynamics systems in equilibrium far from a 
phase transition. For thermodynamic systems in equilibrium but close to a phase 
transition the power spectrum of thermal flctuations is peaked at modes with 
very long wavelength ${\cal P}(k) \sim \zeta^2\ d^{3-\gamma}\ k^{-\gamma}$, with 
$\gamma \in (0, 3)$, and the associated two-points correlation function is 
long-range. When this is the case 

\begin{eqnarray}
\nonumber
\label{variancePhT}
(\Delta M_V)^2 = 8\zeta^2 \rho_0^2\ L^{3+\gamma}\ d^{3-\gamma}\ 
\int_0^{L/d}\ d(k L)\ \frac{1}{(k L)^{\gamma}} \left(\frac{1}{(k L)^2} 
sin(k L) - \frac{1}{k L} cos(k L) \right)^2 \\
\simeq \zeta^2\ \rho_0^2\ d^{3-\gamma}\ V^{1+\gamma/3}.
\end{eqnarray}
The largest contribution to the variance of spatial density anisotropies still comes 
from modes $k L \sim 1$ whose wavelength is comparable to the size of the volume
of integration, but now the variance grows with a higher power of the volume 
of integration. Estimation (\ref{approximation}) still gets it right. 
 
We have brought these examples of thermal anisotropies into consideration 
because we want to notice that their power spectra are not suppressed over any 
range of short momenta, $lim_{k \rightarrow 0} {\cal P}(k) \neq 0$. 
In the statistical systems described by these power spectra there is no causal 
horizon. These systems are theoretically allowed to excite with finite probability 
random density fluctuations with any wavelength longer than the resolution scale 
$d$. In consequence, in these statistical systems the bulk contribution (\ref{C1}) 
to the variance (\ref{twocontributions}) of spatial random anisotropies in 
macroscopic volumes $V$ of any size $L$ is always dominant over the scale 
invariant contribution (\ref{C2}). 

The power spectrum of an statistical system that is constrained by a finite 
causal horizon $D$ looks like 

\begin{equation}
\label{power}
{\cal P}(k) = p(k\ D)\ {\cal P}_0(k \sim D^{-1}), 
\end{equation}
where the dimensionless function $p(k\ D) \simeq o(k\ D)$ decreases to zero 
faster than linearly for $k \ll D^{-1}$ and $p(k\ D) \simeq 1$ for 
$k \simeq D^{-1}$. The pattern of anisotropies associated to these power
spectra are typically of the form (\ref{varianceT}) 
or (\ref{variancePhT}) over volumes of size $L \simlt D$ shorter than the 
horizon, but it turns scale invariant (\ref{scaling}) over volumes of size 
$L \gg D$ as a result of the suppression of the power spectrum (\ref{power}) 
in modes $k \ll D^{-1}$

In the model discussed in subsection II.A, for example, the power spectrum 
(\ref{powerbrownian}) is suppressed in modes $k \ll D^{-1}$ by a factor 
$p(k\ D) \sim (k\ D)^4$. A similar suppression factor appears in the power 
spectrum of cosmological density anisotropies produced at decoupling by 
standard physics operating during the radiation dominated expansion.
A suppression factor $p(x) = \frac{x^n}{1+x^n}$, $n \ge 1$,
appears in the power spectrum that describes the statistical distribution of 
topological defects in a sample cooled down homogeneously, but very fast, below 
its critical temperature (Kibble mechanism) \cite{Rajantie}. The super fast 
cooling rate of the sample, rather than the cosmological expansion, introduces an 
''horizon'' in this example, if we define the horizon $D$ of an homogeneous and 
isotropic statistical system as the scale $D^{-1}$ beyond which its power 
spectrum is suppressed. Spatial random anisotropies in the distribution of 
defects over regions of the sample of size much larger than the "cooling" 
horizon are theoretically known to be scale invariant. 

The suppression of the power spectrum (\ref{power}) over scales $k \ll D^{-1}$ 
introduces the global constrain

\begin{equation}
\label{incompressibility}
\left[\int_{\Omega} d^3{\vec r}\ F(r) = lim_{k \rightarrow 0} 
{\cal P}(k)\right] \hspace{0.3in} \ll \hspace{0.3in} {\cal P}(k \sim D^{-1}).
\end{equation}
This condition $\int_{\Omega} d^3{\vec r}\ F(r) \simeq 0$, which we have 
associated directly to the scale invariance of random anisotropies over very 
large volumes, is known in the literature as a demand of incompressibility of 
the statistical density fluctuations. Incompressibility is understood in our 
context as a constrain $(\Delta M_{\Omega})^2 = 0$ of total conservation of 
the extensive magnitude $M$ over the whole volume $\Omega$, see section IV in 
\cite{oaknin1}. This condition can be evidently satisfied when the sub-volumes $V$ of any size 
are assumed to be embedded in a larger isolated microcanonical ensemble.
In the framework of a quantum field theory this condition can be easily 
obtained if the state of the system is assumed to be an eigenstate of 
$M_{\Omega}$. In quantum Hall effect in condensed matter physics, for 
example, the incompressibilty of the electron fluid is known  to be 
associated to the low-energy border exitations of the system \cite{Halperin}.

\subsection{Amplitude of primordial cosmological density anisotropies.}

In this subsection we examine the amplitude of the scale invariant cosmological 
density anisotropies left at decoupling by standard physics (\ref{scaling}) and 
compare it with the amplitude of the scale invariant anisotropies 
(\ref{scaleinvariance}) imprinted in the CMBR.  

We describe the radiation dominated density field at (physical) cosmological 
time $t \le t_{eq}$ as a statistically homogeneous and isotropic 
random field,

\begin{equation}
\rho({\vec x}; t) = \rho_t + \rho_t \int \frac{d^3{\vec k}}{(2\pi)^3}\
\delta_{t; {\vec k}}\ e^{-i{\vec k}\cdot{\vec x}},
\end{equation}
with 

\begin{equation}
\label{zero_av}
\langle \delta_{t;{\vec k}} \rangle = 0,  \hspace{0.3in}\left(
\langle \rho({\vec x}, t)\rangle = \rho_t\right),
\end{equation}
and

\begin{equation}
\label{power_t}
\langle \delta^*_{t;{\vec k}_1} \delta_{t;{\vec k}_2} \rangle =
(2\pi)^3\ {\cal P}(|{\vec k}_1|; t)\ \delta^3({\vec k}_1 - {\vec k}_2).
\end{equation}
The Fourier modes are assumed to be statistically independent and gaussian. 
We keep using comoving spatial coordinates defined with the scale factor 
of the universe at decoupling $a(t_{eq}) = 1$ as reference. 

The statistical variable $\rho({\vec x}; t) - \rho({\vec y}; t)$
describes the random fluctuations in the density at some comoving
site ${\vec x}$ with respect to a different comoving site ${\vec y}$, at the given 
cosmological time $t$. Statistical homogeneity  (\ref{zero_av}) demands that 
this statistical variable has zero average value, 

\begin{equation}
\langle \rho({\vec x}; t) - \rho({\vec y}; t) \rangle = 0. 
\end{equation}
We are interested in the variance of its statistical fluctuations,
 
\begin{eqnarray}
\label{differenceVAR}
\nonumber
\langle \left[\rho({\vec x}; t) - \rho({\vec y}; t)\right]^2
\rangle & = & 2\left[\langle \left[\rho({\vec x}; t)\right]^2 \rangle -
\langle \rho({\vec x}; t) \rho({\vec y}; t) \rangle\right] = \\
\nonumber
& = & 2\left[\langle \left[\rho({\vec x}; t)\right]^2 \rangle - \rho^2_{t}\right]
- 2\left[\langle \rho({\vec x}; t) \rho({\vec y}; t) \rangle - 
\rho^2_{t}\right] = \\
\nonumber 
& = & 2\rho^2_{t} \int \frac{d^3{\vec k}}{(2\pi)^3} {\cal P}(k; t) 
\left(1 - e^{+i{\vec k}\cdot({\vec x}-{\vec y})}\right) = \\
\label{differenceVAR}
& = & \frac{\rho^2_{t}}{\pi^2} \int dk\ k^2\ {\cal P}(k; t) \left(1 - 
\frac{Sin(|{\vec k}||{\vec x}-{\vec y}|)}{|{\vec k}||{\vec x}-{\vec y}|}\right). 
\end{eqnarray} 

During the radiation dominated expansion of the universe the cosmic
fluid supports the propagation of non-linear density 
waves, which travel with sound velocity $c_s = 1/\sqrt{3}$ 
(in natural units) and operate to erase density inhomogeneities. 
Therefore, we must expect maximal statistical correlation between causally 
connected comoving sites:

\begin{eqnarray}
\label{differenceVAR1}
\langle \left[\rho({\vec x}; t) - \rho({\vec y}; t)\right]^2
\rangle \ll 2\left[\langle \left[\rho({\vec x}; t)\right]^2 
\rangle - \rho^2_{t}\right],
\hspace{0.3in} for \hspace{0.3in} |{\vec x}-{\vec y}| \simlt H^{-1}_{t}.
\end{eqnarray}
On the other side, causality constrains forbide the existence of 
statistical correlation between causally disconnected sites:

\begin{eqnarray}
\label{differenceVAR2}
\langle \left[\rho({\vec x}; t) - \rho({\vec y}; t)\right]^2
\rangle \simeq 2\left[\langle \left[\rho({\vec x}; t)\right]^2 
\rangle - \rho^2_{t}\right],
\hspace{0.3in} for \hspace{0.3in} |{\vec x}-{\vec y}| \gg H^{-1}_{t}.
\end{eqnarray}
Actually, causality does constrain the two points correlation function to 
satisfy the more stringent condition (\ref{incompressibility}). 
The comoving causal horizon $H^{-1}_t = (t/t_{eq})^{1/2}\ H^{-1}_{eq}$ 
grows monotonically during the radiation dominated expansion of the universe.

These features correspond to a power spectrum ${\cal P}(k; t)$ peaked at 
$k \sim H_{t}$, 

\begin{equation}
\label{peak}
{\cal P}(k \sim H_{t}; t) \equiv {\cal P}_{t}, 
\end{equation}
and suppressed both at $k \gg H_{t}$ and at $k \ll H_{t}$,

\begin{eqnarray}
\label{FRWpower}
{\cal P}(k; t) \le \left\{ 
\begin{array}{c}
{\cal P}_{t}\  H^3_{t}/k^3, \hspace{0.3in} for \hspace{0.3in} 
k \gg H_{t}, \\
{\cal P}_{t}\ k^4/H^4_{t}, \hspace{0.3in} for \hspace{0.3in} 
k \ll H_{t},
\end{array}
\right.
\end{eqnarray}
We plug this power spectrum in expression (\ref{variance}) 
to obtain the variance of the density anisotropies left in the 
cosmic plasma at cosmological time $t$: \\
 
$\left. I\left. \right. \right)$ The suppression of the power in
modes $k \gg H_t$ implies that over volumes $V$ of comoving size 
$L \simlt H^{-1}_{t}$ smaller than the causal horizon the anisotropies
are extensive,

\begin{equation}
\label{within}
(\Delta M_V)^2(t) \simeq \rho^2_{t} \left(H^3_{t} {\cal P}_{t}\right)\ V^2
\simeq (\Delta M_{V_{t}})^2 \frac{V^2}{V^2_{t}}. 
\end{equation} 
In particular, over comoving volumes $V_t$ of size 
$L \sim H^{-1}_t$ comparable to the causal horizon

\begin{equation}
\label{amplB}
(\Delta M_{V_{t}})^2(t) \simeq \left(\rho^2_{t} H^{3}_{t} 
{\cal P}_{t}\right)\ V^2_{t} \simeq \rho^2_{t} H^{-3}_{t} {\cal P}_{t}.
\end{equation} 

$\left. II \right)$ The suppression of the power in modes $k \ll H_t$ 
implies that over volumes $V$ of comoving size $L \gg H^{-1}_{t}$ much larger 
than the causal horizon the anisotropies are scale invariant

\begin{equation}
\label{outside}
(\Delta M_V)^2(t) \simeq \rho^2_{t} H^{-1}_{t} {\cal P}_{t}\ L^2
\simeq (\Delta M_{V_{t}})^2 \frac{L^2}{H^{-2}_{t}}.
\end{equation}

The amplitude of the cosmological density anisotropies (\ref{within}), 
(\ref{outside}) at decoupling $t = t_{eq}$ is commensurate with the amplitude 
of the primordial anisotropies imprinted in the CMBR if, and only if,
condition (\ref{Peebles}) is fulfilled. From (\ref{amplB}) we obtain the
condition $H^3_{eq}\ {\cal P}_{t_{eq}} \sim 10^{-10}$. 
In order to give it a clear physical meaning we now define

\begin{equation}
\label{parameter}
\kappa^2_t \equiv \frac{1}{\rho^2_t} \left[\langle \left[\rho({\vec x}; t)
\right]^2 \rangle - \rho^2_{t}\right] =
\int \frac{d^3 {\vec k}}{(2\pi)^3}\ {\cal P}(k; t) \simeq 
H^3_t\ {\cal P}_t, 
\end{equation}
and notice that it is the normalized variance of the density fluctuations between
comoving sites ${\vec x}$ and ${\vec y}$ that are causally disconnected at 
time $t$,

\begin{equation}
\label{parameterB}
\kappa^2_t \simeq \frac{1}{2} \frac{1}{\ \rho^2_t} \langle \left[\rho({\vec x}; t) - 
\rho({\vec y}; t)\right]^2 \rangle,
\hspace{0.3in} for \hspace{0.3in} |{\vec x}-{\vec y}| \gg H^{-1}_{t}. 
\end{equation}
We have from (\ref{amplB}) and (\ref{parameter})

\begin{equation}
\label{thekey}
(\Delta M_{V_{t}})^2(t) \simeq \kappa^2_t\ \rho^2_t\ V^2_{t}. 
\end{equation}
Hence, condition (\ref{Peebles}) simply requires

\begin{equation}
\label{condition_kappa}
\kappa^2_{t_{eq}} \simeq 10^{-10}.
\end{equation}
As there does not seem to exist any obvious theoretical or observational 
constraint that rules out (\ref{condition_kappa}) in the framework of standard 
FRW cosmology, we should conclude from this analysis that primordial 
cosmological structures (\ref{scaleinvariance}) could have been seeded during 
the radiation dominated expansion of the universe. 

In a forthcoming paper we will explore the mechanisms involved
in actually fixing $\kappa_t$ at the time $t = t_{eq}$ of decoupling.
We will show that $\kappa^2_t$  is a measure of the entropy of the random 
density fluctuations in the cosmic plasma at time $t$.
This observation is important if the entropy of the density fluctuations
is preserved during the adiabatic radiation dominated expansion of the 
universe. In such a case, $\kappa^2_t$ is constant since the much earlier 
time $t_*$ when the universe entered the stage of adiabatic expansion,

\begin{equation}
\label{adiabaticity}
\kappa^2_t \simeq constant, \hspace{0.3in}  for \hspace{0.3in} 
t \in (t_*, t_{eq}).
\end{equation}
This would imply that the amplitude of the anisotropies (\ref{within}), 
(\ref{amplB}), (\ref{outside}) at decoupling $t = t_{eq}$ was indeed fixed by 
entropy-producing mechanisms at that earlier time $t_*$. 
Notice that condition (\ref{adiabaticity}) is satisfied if the relative 
amplitude of the density fluctuations between any two causally disconnected 
sites (\ref{parameterB}) is preserved during the adiabatic expansion.

\section{Conclusions.}

We have shown that the prevalent analysis of the origin of primordial 
structures in standard FRW cosmology is flawed by an unjustified approximation, 
which has led researchers to erroneous conclusions. 
Our revised analysis of the issue shows that standard physics operating in the
cosmic plasma during the radiation dominated expansion of the universe can
produce scale invariant cosmological density anisotropies at decoupling with
amplitude comparable to that of the primordial anisotropies imprinted in
the CMBR. Indeed, the characteristic scale invariance of the primordial 
cosmological density anisotropies at decoupling seems to be a signature 
of the causal structure of the standard FRW universe, which prevents the 
development in the radiation dominated cosmic plasma of large density 
fluctuations with cosmologically large comoving wavelength. This might explain
the reported absence of statistical correlations at large angular scales in the 
temperature anisotropies measured by the WMAP \cite{Copi:2006tu}. 

{\bf Acknowledgments}
I wish to thank my teacher A. Oaknin~(z.l). This work is the result of
her example of honesty and dedication.
I also wish to thank Prof. R.~Brustein and Prof. A.~Vilenkin 
for their useful comments.

\end{document}